\documentclass[aps,superscriptaddress,showpacs,twocolumn]{revtex4}
\usepackage[dvips]{graphicx}
\usepackage{amssymb}

\def \beq {\begin{equation}}
\def \eeq {\end{equation}}

\begin{document}

\title{Destroying a near-extremal Kerr-Newman black hole}

\author{Alberto Saa}\email{asaa@ime.unicamp.br}
\affiliation{Departamento de Matem\'atica Aplicada, UNICAMP,  13083-859 Campinas, SP, Brazil}
\author{Raphael Santarelli}\email{telsanta@ifi.unicamp.br}
\affiliation{Instituto de F\'\i sica ``Gleb Wataghin'', UNICAMP,
  13083-859 Campinas, SP, Brazil}

\begin{abstract}
We revisit here a previous  argument due to Wald showing  the
impossibility of turning an extremal Kerr-Newman black hole into a naked
singularity by plunging test particles across the black hole event
horizon. We extend Wald's analysis to the case of   near-extremal black holes
and show that it is indeed possible to destroy their event horizon, giving rise to   naked singularities, by pushing test particles towards the
black hole as, in fact, it has been demonstrated explicitly
by several recent works.
Our analysis allows us   to go a step further and
 to determine the optimal values,
in the sense of keeping
to a minimum the backreaction effects,  of the
  test particle
 electrical charge and angular momentum  necessary  to destroy a given near-extremal Kerr-Newman black hole.
 We describe briefly a
 possible realistic scenario for
 the creation of a Kerr naked singularity
 from some recently discovered candidates to be rapidly rotating black holes in  radio galaxies.
\end{abstract}

\maketitle

\section{Introduction}

There has been recently a revival of interest in the problem of turning
a black hole   into a naked singularity by means of classical
and quantum processes,
see, for instance, \cite{Matsas:2007bj} and \cite{Jacobson:2010iu} for references and a brief review with a historical perspective. Such a
 problem is intimately related
to the weak cosmic censorship conjecture\cite{Wald:1997wa,Clarke}. Indeed, the typical facility in covering a naked singularity with an
 event horizon  and the apparent
impossibility of destroying a black hole horizon \cite{Cohen:1979zzb} have strongly endorsed  the
validity of the conjecture along the years, albeit
it has started to be challenged
recently.

The most generic asymptotically flat black hole solution of Einstein equations   we can consider is the Kerr-Newman black hole,
which is completely
characterized by its mass $M$, electric charge $Q$ and angular momentum $J=aM$. The distinctive feature of a black hole, namely the existence of an event horizon covering the central singularity, requires
\beq
\label{cond}
M^2 \ge a^2 + Q^2,
\eeq
with the
  equality corresponding  to the so-called extremal case. If (\ref{cond})
does not hold, the central singularity is exposed, giving rise to a naked spacetime
 singularity, which should not exist in Nature according to the
weak cosmic censorship conjecture.
All the classical results on the impossibility of destroying black hole horizons were obtained by considering extremal black holes. The first
work arguing  that it would be indeed possible to destroy the horizon of a near-extremal black hole is quite recent and it was due to Hubeny \cite{Hubeny:1998ga}, which considered a Reissner-Nordstr\"om $(a=0)$ black hole. The physical possibility
of destroying the horizon of a near-extremal black hole by over-spinning
or over-charging it with the absorption  of test particles or fields is
nowadays a very active field of research and
debate\cite{Matsas:2007bj,Jacobson:2010iu,Richartz:2008xm,Hod:2008zza,Matsas:2009ww,Jacobson:2009kt,Berti:2009bk,BouhmadiLopez:2010vc,Jacobson:2010it,Chirco:2010rq,Barausse:2010ka,Rocha:2011wp}.

The impossibility of destroying the event horizon  by plunging  test particles into   an extremal Kerr-Newman black hole
 is clear and elegantly  summarized in Wald argument \cite{Wald}, which we briefly  reproduce here. We consider test particles
 with energy $E$, electric charge $e$, and orbital angular momentum $L$.
  The test particle approximation     requires  $E/M \ll 1$, $L/aM \ll 1$,
  and $e/Q \ll 1$,  assuring, in this way, that   backreaction effects
  are negligible.
The particle angular momentum $L$ is assumed to be aligned with the black hole angular momentum $J$, and both $Q$ and $e$ are assumed, without loss of generality, to be positive.   According to the laws of black hole thermodynamics (see, for instance, section 33.8 of \cite{MTW}),
after the capture of a test particle, the black hole
will have total angular momentum $aM + L$, charge $Q+e$ and mass no greater than $M+E$. In order to form a naked singularity, one needs
\beq
(M+E)^2 < \left(\frac{aM + L}{M+E} \right)^2 + (e+Q)^2,
\eeq
which implies, for extremal black holes and in the test particle approximation,
\beq
\label{en}
E < \frac{QeM + aL}{M^2+a^2}.
\eeq
However, in order to assure that the test particle be indeed plunged into the black hole, its energy must obey\cite{MTW}
\beq
\label{min}
E \ge E_{\rm min} =
\frac{Qer_+ + aL}{r_+^2 + a^2},
\eeq
with
\beq
r_+ = M + \sqrt{M^2- a^2 -Q^2}
\eeq
being the event horizon radius of the black hole. For an extremal black hole,
$r_+ = M$, and it is clear that (\ref{en}) and (\ref{min}) will not be
fulfilled simultaneously, implying that one cannot turn an extremal
Kerr-Newman black hole into a naked singularity by plunging test particle
across its event horizon. Wald presents also a similar argument for the case of dropping  spinning uncharged particles into a Kerr ($Q=0$) black hole. Notwithstanding, de Felice and Yunqiang\cite{deFelice:2001wj} showed that it would be indeed possible to transform a
Reissner-Nordstr\"om black hole in a Kerr-Newman naked singularity after capturing an electrically neutral spinning body.

The purpose of this work is to extend Wald's original
analysis \cite{Wald} to the case
of near-extremal Kerr-Newman black holes and show explicitly that it is indeed possible to over-spin and/or over-charge
near-extremal black holes
  by plunging test particle across their event horizon while  keeping
  backreaction effects to a minimum. All the recently proposed
mechanisms to destroy a near-extremal black-hole by using infalling test particles are accommodated in our analysis. Furthermore,
we  determine the
optimal values, in the sense that they keep backreaction effects to
a minimum,
of the electrical charge and angular momentum of the incident test particle in order to destroy a   near-extremal Kerr-Newman black hole with given mass, charge, and angular momentum.  We show also that it is not strictly  necessary to
plunge the particles across the black hole horizon, but they can be thrown
from infinity and proceed towards to the black hole following a geodesics, minimizing in this way any back reaction effect associated
with the specific mechanism to release the particle near, os push it
against, the black hole horizon. As an explicit example, we consider some recently
discovered candidates to be rapidly rotating black holes in  radio galaxies
and show how it would be possible to create Kerr naked singularities from them
with minimal backreaction effects.

\section{Near-extremal Kerr-Newman black holes}
We call near-extremal a Kerr-Newman black hole for which
\beq
\delta^2 = M^2 - a^2 - Q^2 > 0, \quad \frac{\delta}{M} \ll 1.
\eeq
For a near-extremal black hole, the condition (\ref{cond}) for the creation of a naked singularity by absorbing a test particle with
energy $E$, electric charge $e$, and orbital angular momentum $L$
implies that
\beq
\label{max1}
E < E_{\rm max} = \frac{QeM + aL}{M^2+a^2} - \frac{M^3}{2(M^2+a^2)}
\left(\frac{\delta}{M} \right)^2.
\eeq
It is more convenient here to introduce a
  parametrization for near-extremal black holes
\begin{eqnarray}
a &=& \sqrt{M^2-\delta^2}\cos\alpha,\\
Q &=& \sqrt{M^2-\delta^2}\sin\alpha,
\end{eqnarray}
with $0\le \alpha \le \pi/2$. In this way,
near-extremal black holes are   characterized   by the triple $(M,\delta,\alpha)$. For instance, near-extremal Kerr and Reissner-Nordstr\"om black holes correspond, respectively, to $(M,\delta,0)$ and $(M,\delta,\pi/2)$. After some straightforward algebra, it is possible to show
that
\beq
  E_{\rm max} = A
 - \frac{  M+A\sin^2\alpha}{2+2\cos^2\alpha}  \left(\frac{\delta}{M} \right)^2,
\eeq
where only terms up to second order in $(\delta/M)$ were kept, and
\beq
A = \frac{(L/M)\cos\alpha + e\sin\alpha}{1+\cos^2\alpha} \ge 0.
\eeq

\begin{figure}[tb]
\includegraphics[width=1\linewidth]{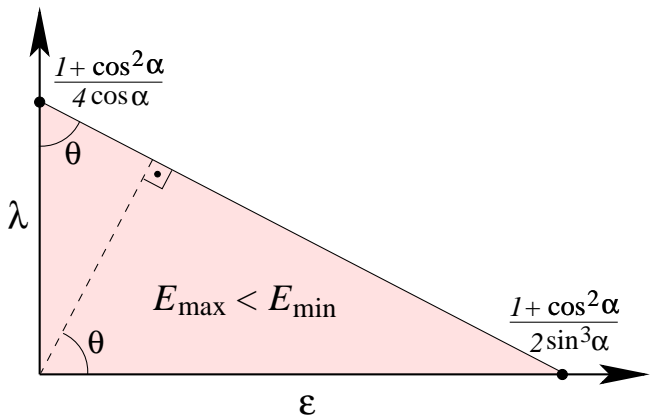}
\caption{In the shadowed region, no naked singularity is formed.
Conditions (\ref{max1}) and (\ref{min1}) are simultaneously
fulfilled in the region above the straight line in the
$(\lambda,\varepsilon)$ plane, where
$\lambda = L/ M\delta$ and $\varepsilon= e/\delta$. The validity of
the test particle approximation, of course, does not allow arbitrary
large values of $\lambda$ and $\varepsilon$. Hence, the allowed region for
the creation of naked singularities is delimited by the
straight line and by the validity of the test particle approximation.}
\label{fig1}
\end{figure}

The event horizon for near-extremal black holes are located at
$r_+ = M + \delta$, which implies   that  the  condition (\ref{min}) assuring that the particle is to be captured reads
\beq
\label{min1}
E \ge E_{\rm min} = A - B\left(\frac{\delta}{M} \right) -
\left(\frac{(2+\sin^2\alpha)A - 4B}{2+2\cos^2\alpha} \right) \left(\frac{\delta}{M} \right) ^2,
\eeq
where, again, only   terms up to $(\delta/M)^2$ were kept and
\beq
B = \frac{2(L/M)\cos\alpha + e\sin^3\alpha}{(1+\cos^2\alpha)^2}\ge 0.
\eeq
It is clear that for the extremal case ($\delta=0$) we have Wald's result
$E_{\rm max}=E_{\rm min}$, implying that (\ref{max1}) and (\ref{min1})
cannot be fulfilled simultaneously. However, for $\delta > 0$ it is
 indeed possible to a test particle obey (\ref{max1}) and (\ref{min1}).
The intersection of $E_{\rm max}$ and $E_{\rm min}$ in the $(\lambda,\varepsilon)$ plane corresponds to
the straight line
\beq
\label{cros}
2\lambda\cos\alpha + \varepsilon\sin^3\alpha = \frac{1+\cos^2\alpha}{2},
\eeq
where
$\lambda = L/ M\delta$ and $\varepsilon= e/\delta$. The forbidden region,
 where no naked singularity is formed, is depicted in Fig. \ref{fig1}.

\subsection{Optimal test particles}

In all   derivations done so far, we have used the test particle approximation.
The idea of minimizing  the values of $E$, $L$,
  and $e$ necessary to destroy the black hole is more than a simple
requirement of consistence. It helps to assure that   back reaction
effects are negligible and, consequently,  that
it will not be possible to restore  the black hole event horizon by means of any
subdominant physical process. The first, and maybe the more natural,
criterium of optimality we can devise here is to require minimal
test particle total energy $E$, which is given by $E_{\rm min}$ in
(\ref{min1}). The test particle
minimal energy $E(\lambda,\varepsilon)$ necessary  to destroy the black hole corresponds to the minimum value of $E_{\rm min}$, subject to the restriction
(\ref{cros}). This is a simple linear optimization problem \cite{opt}, and
the solution is known to correspond to one of the points $(0,\varepsilon)$
or $(\lambda,0)$, {\em i.e.}, the  minimal energy $E$ will be
given either by
\beq
\label{e1}
E(0,\varepsilon) = \frac{\delta}{2\sin^2\alpha}
\eeq
or
\beq
\label{e2}
E(\lambda,0) = \frac{\delta}{4} .
\eeq
It is clear that $E(\lambda,0) < E(0,\varepsilon)$ for any value of
$\alpha$, suggesting that the best option to turn a black hole into
a naked singularity would be to plunge an uncharged particle, irrespective of
the value of $\alpha$. However, we see from (\ref{cros}) that
$\lambda$ can increase considerably for small $\alpha$, despite of
$E(\lambda,0)$ being a minimum. The minimization of $E(\lambda,\varepsilon)$ does not guarantee the minimization of $L$ and $e$, risking the validity
of the test particle approximation. In order to avoid theses problems,
we will require that $\lambda^2 + \varepsilon^2$ be minimal
for optimal test particles.
Such a requirement
 corresponds to select the  nearest point
of the straight line (\ref{cros}) to the origin in the $(\lambda,\varepsilon)$ plane. From Fig. \ref{fig1} and simple trigonometry, we
have that a test particle with angular momentum $L$ and electrical
charge $e$ obeying
\beq
\label{opt}
 \frac{eM}{L} = \frac{\sin^3\alpha}{2\cos\alpha} =
 \frac{(Q/a)^3}{2+2(Q/a)^2}
\eeq
is the optimal test particle to turn a near-extremal Kerr-Newman black with parameters $(M,\delta,\alpha)$ into a naked singularity. The energy associated with the optimal parameters (\ref{opt}) is in the range
delimited by (\ref{e1}) and (\ref{e2}).

It is clear from (\ref{opt}) that the optimal test particle to turn a
near-extremal Kerr black hole $(\alpha=0)$ into a naked singularity should be electrically neutral $(e=0)$. Moreover, from Fig. \ref{fig1}, we see
that the particle must have $L/M\ge\delta/2$. The minimum particle
 energy is given by (\ref{e2}), assuring the validity of the
 test particle approximation in this case
  ($E/M \ll 1$ and $L/aM \ll 1$) provided the black hole be near-extremal
 ($\delta/M \ll 1$).  For the case of a Reissner-Nordstr\"om black hole
 $(\alpha=\pi/2)$, the optimal test particle must be charged, with  $e>\delta/2$, and have vanishing orbital angular momentum
 ($L=0$). The  minimum particle
 energy for this case is given by $E=\delta/2$, according to (\ref{e1}).
 The validity of the test particle approximation and the minimization of any backreaction effect is assured also in this case.

\section{Throwing particles from infinity}

The expression for $E_{\rm min}$ given by equation (\ref{min}) corresponds to the minimal energy that a test particle can have at the black hole horizon.
On the other hand, the minimal energy that a particle can have anywhere on
the equatorial plane outside a Kerr-Newman black hole is given by the effective potential \cite{MTW}
\beq
\label{pot}
V(r) = \frac{\beta + \sqrt{\beta^2 -\nu\gamma_0} }{\nu},
\eeq
where
\begin{eqnarray}
\label{b1}
\beta &=& (La + Qer)(r^2+a^2) - La\Delta,\\
\label{b2}
\nu  &=& (r^2+a^2)^2 -  a^2\Delta,\\
\label{b3}
\gamma_0  &=& (La + Qer)^2 - ( L^2 + \mu^2r^2)\Delta,
\end{eqnarray}
with $\mu$ being the particle rest mass and
\beq
\Delta = r^2 - 2Mr + a^2 + Q^2.
\eeq
The event horizon $r_+$ is the outermost zero of $\Delta$ and
equation $(\ref{min})$ corresponds to the potential (\ref{pot})
evaluated on $r=r_+$. For $r\to \infty$ one has $V\to \mu$, as it is
expected for any asymptotically flat solution. For a particle of energy
$E$, the points for which $V(r)=E$ are return points and delimit the
classically allowable region for the particle motion. In order to assure
that a particle with energy $E$ thrown from infinite reaches the horizon, we need to have
$E>V(r)$ in the exterior region of the black hole. In some cases, one
can have that $\mu > E_{\rm min}$, {\em i.e.}, the energy necessary for the
incident test particle reach the horizon
 is smaller than its rest mass. This is not a surprise in gravitational systems, but, of course, this trajectory cannot start from infinity.
To follow
  this trajectory, a particle must be released near the horizon, by
some external mechanism. This is an extra and unnecessary complication
in our analysis. The external mechanism could be subject to some
backreaction or subdominant physical effect that could eventually prevent
the particle of entering the black hole. This can be avoided if we
adjust the particle rest mass $\mu$ properly. An explicit example can
enlighten this point.

Let us consider the case of a Kerr black hole $(Q=0)$. (For a
recent review on   equatorial orbits in   Kerr black holes and
naked singularities, see
\cite{Pugliese:2011xn}.) For
a near-extremal Kerr black hole, the effective potential (\ref{pot})
can be written as
\beq
V(r) = \tilde{V}(r) + O\left(\left(\delta/M\right)^2\right),
\eeq
where $\tilde{V}(r)$
stands for the effective potential for the extremal Kerr
black hole, which can be calculated from (\ref{pot}) with
$Q=0$ and $a=M$.
\begin{figure}[tb]
\includegraphics[width=1\linewidth]{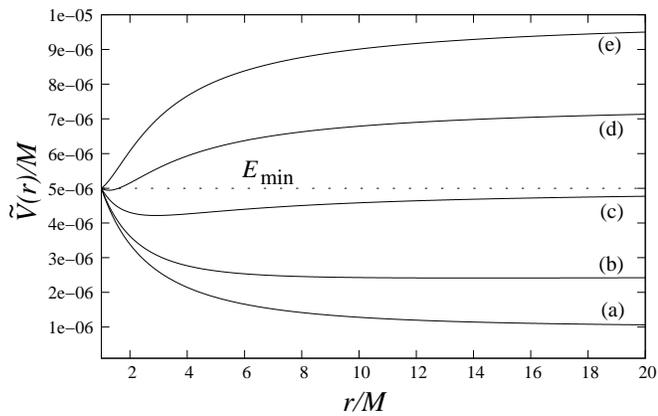}
\caption{The effective potential (\ref{pot}) for a test particle
with angular momentum $L$ and rest mass $\mu$ around an extremal
Kerr black hole with mass $M$. For this figure, $L/M^2=10^{-5}$ and
$\mu/M = 10^{-6}$(a), $2.5\times 10^{-6}$(b), $5\times 10^{-6}$(c),
$7.5\times 10^{-6}$(d), and $10^{-5}$(e). The doted horizontal line corresponds to
the value  of the effective potential at the black hole horizon ($E_{\rm min}$), which does not dependent on $\mu$.}
\label{fig2}
\end{figure}
 Fig. \ref{fig2} depicts the
effective potential  $\tilde{V}(r)$ for different values of
$\mu$. Typically, the choice of $\mu < E_{\rm min}$ will allow the
particle to reach the horizon when thrown  from infinite with
energy $E\approx E_{\rm min}$.

\section{Final remarks}

We have shown that the Wald analysis \cite{Wald} can be extended to
the case of near-extremal Kerr-Newman black holes, allowing the
accommodation of the recent proposals to over-spin or over-charge
near-extremal black holes in a single and simpler framework. Moreover, we could
determine the optimal parameters for a test particle in order to
destroy a black
hole while keeping backreaction effects to a minimum. An explicit and
realistic example
here will be valuable  to enlighten these points.

There are some evidences of rapidly rotating black holes in quasars
\cite{Wang:2006bz} and radio galaxies \cite{Wu:2011wz}. These black holes
are very massive, having
 typically  $M\approx 10^8M_\odot$, and  can attain an angular
 momentum such that $a/M \gtrsim 0.9$.
Let us suppose we have one of these black holes with $\delta/M\approx 10^{-5}$. According to Section 2, the capture of a test body with $L/M^2 =  10^{-5}$ will suffice to the creation of a naked singularity. For this
case, $E_{\rm min} \approx 10^{3}M_\odot$. A test body with
 mass comparable to the moon mass,
$\mu \approx 4\times 10^{-8}M_\odot$,
will certainly be able to reach the horizon if thrown from infinity with
angular momentum $L/M^2 =  10^{-5}$ (See Fig. 2).
The minimal necessary energy for a test body reach the horizon with such
orbital angular momentum is given by (\ref{min1}),
$E_{\rm min}/M = 10^{-5}(1-10^{-5})/2$. On the other hand, with this angular
momentum, any test body captured with energy $E/M < E_{\rm max}/M =
10^{-5}(1-10^{-5}/2)/2$ will destroy the black hole.
Hence, any test body thrown from infinity with
angular momentum $L/M^2 =  10^{-5}$ and energy $E$ such that
$E_{\rm min} < E < E_{\rm max}$ will produce a naked singularity.
Furthermore,
the validity of the test particle approximation is assured in this
example.
 It is important also to notice that
the horizon radius of such rapidly rotation black holes are of the order of
$10^8$ km, very large when compared with the moon radius of
$1.7\times 10^3$ km. A body with mass and size comparable the the moon would
be well described by the test particle approximation even when crossing
the horizon of such rapidly rotating black holes. Indeed,
it is very hard to
devise any backreaction effect that could prevent the formation of a
naked singularity in this case. We close noticing that, probably, the
Thorne limit $a/M\approx 0.998$ \cite{Thorne} corresponds to the most
realistic near-extremal astrophysical Kerr black hole. For such case, $\delta/M\approx 6\%$ and we have analogous results to the preceding example. However, in
this case, the validity of the test particle approximation could be questioned and we could have appreciable backreaction effects.

\acknowledgements

This work was supported by FAPESP and CNPq. The authors wish to thank
V.E. Hubeny for enlightening conversations.

\end{document}